\documentclass[twocolumn,showpacs,prl]{revtex4}
\usepackage{graphicx}
\usepackage{dcolumn}
\usepackage{bm}

\begin{document}

\title{Relaxation Mechanism for Ordered Magnetic Materials}

\author{C. Vittoria}
\author{S.D. Yoon}
\affiliation{Center for Microwave Magnetic Materials and Integrated Circuits\\  
ECE Department, Northeastern University, Boston MA. 02115 USA}
\author{A. Widom}
\affiliation{Physics Department, Northeastern University, Boston MA. 02115 USA}

\begin{abstract}
We have formulated a relaxation mechanism for ferrites and ferromagnetic metals 
whereby the coupling between the magnetic motion and lattice is based purely on 
continuum arguments concerning magnetostriction. This theoretical approach contrasts 
with previous mechanisms based on microscopic formulations of spin-phonon interactions 
employing a discrete lattice. Our model explains for the first time the scaling of the
intrinsic FMR linewidth with frequency, and \begin{math} \frac{1}{M}  \end{math} temperature
dependence and the anisotropic nature of magnetic relaxation in ordered magnetic materials,
where \begin{math} M  \end{math} is the magnetization. Without introducing adjustable 
parameters our model is in reasonable quantitative agreement with experimental 
measurements of the intrinsic magnetic resonance linewidths of important class 
of ordered magnetic materials, insulator or metals.    
\end{abstract}

\maketitle

\section{Introduction \label{intro}}
Since the discovery of magnetic resonance, the physics community has been fascinated with 
possible mechanisms to explain the absorption linewidth or the relaxation time in magnetic 
materials. It was and still is a very challenging problem. Magnetic relaxation is so important 
to understand because it affects a number of technologies, including computer, microwave, 
electronics, nanotechnology, medical, etc.. Ultimately, the physical limitation of any 
technology which incorporates magnetic materials of any size, shape and combinations thereof 
comes down to precise knowledge of the relaxation time of the magnetic material being utilized. 
The background of various calculations or formulations of magnetic relaxation 
for the past sixty years or so can be summarized briefly as follows: (i) The relaxation times 
in paramagnetic materials \cite{Bloch:1932} is characterized by two parameters, 
\begin{math} T_1  \end{math} and \begin{math} T_2  \end{math},
wherein \begin{math} T_2^{-1}  \end{math} describes the magnetic resonance linewidth and  
\begin{math} T_1  \end{math} describes the time taken for the external magnetic field 
Zeemann energy density \begin{math} -{\bf H}_{ext}\cdot {\bf M}  \end{math} to relax into 
thermal equilibrium. These times have been modeled in terms of various coupling schemes, 
i.e.  spin-spin and/or spin-lattice interactions \cite{Callen:1962}. Since the coupling between spins 
is relatively weak, as it should be in a paramagnetic material, the coupling to the lattice involves 
discrete spin sites rather than a collective cluster of spins. As such, paramagnetic coupling is 
necessarily microscopic in nature. For example, a microscopic coupling scheme was 
formulated \cite{Orbach:1961} whereby a spin Hamiltonian was modulated by the lattice motion. 
Variants to this approach have been very successful in explaining relaxation in paramagnetic 
materials. (ii) The magnetic relaxation of ferrimagnetic or ferromagnetic resonance (FMR) linewidth 
is characterized by the Gilbert parameter \cite{Gilbert:2004} \begin{math} \alpha  \end{math},   
or equivalently by Landau-Lifshitz parameter \cite{Landau:1935} \begin{math} \lambda_L  \end{math}. 
A distinguishing feature of the collective coherent magnetic moments in FMR is that the magnitude 
of the magnetization, \begin{math} M=|{\bf M}| \end{math} remains fixed which requires a magnetic 
resonance equation of the simple form 
\begin{equation}
\frac{d{\bf M}}{dt}=\gamma {\bf M}\times {\bf H}_{tot}=
\gamma {\bf M}\times ({\bf H}+{\bf H}^\prime ),
\label{intro1}
\end{equation} 
wherein the gryomagnetic ratio \begin{math} \gamma =ge/2mc  \end{math}. The {\em total} magnetic 
intensity \begin{math} {\bf H}_{tot} \end{math}
has a thermodynamic part \begin{math} {\bf H} \end{math} determined by the energy per unit 
volume \begin{math} u \end{math}, 
\begin{equation}
du=Tds+{\bf H}\cdot d{\bf M},
\label{intro2}
\end{equation} 
and a dissipative part \begin{math} {\bf H}^\prime \end{math} determined by the Gilbert linear 
operator 
\begin{math} \tilde{\alpha }  \end{math},
\begin{equation}
{\bf H}^\prime =\left[\frac{1}{\gamma M}\right]
\tilde{\alpha}\cdot \frac{d{\bf M}}{dt}\ .
\label{intro3}
\end{equation} 
Eqs.(\ref{intro1}) and (\ref{intro3}) imply that all components of the magnetization must relax 
simultaneously in a way which conserves the magnitude of the magnetization. Much of the successful 
microscopic approaches or formulations utilized in paramagnetic materials were transferred over to 
models \cite{Kittel:1962} which attempted to explain Eqs.(\ref{intro1}) and (\ref{intro3}). In some 
sense this presented a contradiction or paradox which was conveniently ignored. As it is well 
known that collective excitations in a ferri or ferromagnetic state can be adequately described in 
classical continuum terminologies, although microscopic descriptions remain perhaps more 
accurate \cite{Sparks:1970}. To our knowledge very few or any microscopic models have been successful 
in explaining the origin of Eq.(\ref{intro3}). 
For example, much attention was given in the seventies 
to explain the FMR linewidth in YIG \begin{math} ({\rm Y_3Fe_5O_{12}}) \end{math}, since 
its linewidth was the narrowest ever measured in a ferrimagnetic material \cite{LeCraw:1962}. 
Clearly, there was less to explain, and perhaps spin-lattice interactions could be treated 
at discrete spin sites as in paramagnetic materials. These calculations \cite{LeCraw:1962} 
contained many approximations and predicted an FMR linewidth about 1/10 to 1/100 
of the measured linewidth. We believe that this is the best agreement between theory 
and experiment on relaxation in an ordered magnetic material. The purpose of this work 
is to improve upon the predictability of a theoretical model not only on a given material 
but in general for any ordered magnetic materials without restoring to any approximations and
assumptions.

We have adopted a conventional continuum magneto-mechanical description of the magnetic and elastic 
states of the ferri or ferromagnetic crystal \cite{Chikazumi:1964,Landau:2000}. The advantage of 
this description is that the microscopic spin-lattice coupling need not be formulated, since it 
has already been included in the continuum model which has been proved to be experimentally correct. 
We introduce a thermodynamic argument stating that the heat exchange between the magnetic and elastic 
systems must be the same. As such, Eq.(\ref{intro3}) may be directly related to the elastic sound 
wave relaxation time and the coupling strength between the magnetic and elastic systems. Specifically, 
we will show that \begin{math} \alpha   \end{math} is proportional to the square of the magnetostriction 
constant. i.e. \begin{math} \lambda^2 \end{math} and inversely proportional to 
\begin{math} \gamma M  \tau \end{math} wherein \begin{math} \tau   \end{math} the elastic relaxation 
time. In addition, the model predicts that \begin{math} \tilde{\alpha} \end{math} cannot be presumed 
to be a scalar as it has been done in the past; i.e. \begin{math} \tilde{\alpha} \end{math} is predicted 
to be anisotropic a second rank tensor in a single crystal material. 

It is clear that one needs an interaction between phonons and electron spins to account for Gilbert damping 
parameter $\alpha $. Suhl \cite{Suhl:1998} and more recently Hickey 
and Moodera \cite{Hickey:2009} have considered such coupling schemes. 
The Gilbert damping parameter can be thought of as a transport coefficient in much the same way as 
conductivity and/or viscosity are transport coefficients. Such transport coefficients describe heating 
processes by which otherwise long lived modes are damped. One can in fact relate the Gilbert damping 
parameter to conductivity and/or viscosity. For metallic ferromagnetic materials, conductivity as well 
as electron viscosity produces considerable amount of magnetic damping via eddy current heating. For magnetic 
insulators it is the viscosity which determines the magnetic damping. As it is well known, conductivity and 
viscosity can be non zero even in zero frequency limit. Hence, the implied Gilbert damping parameter is also 
non zero at zero frequency. In Suhl and Hickey and Moodera's papers they find, in the limit of zero frequency 
and zero wave number, that the real part of $\alpha $ is zero. This limiting case suggests that they have not 
included the zero frequency transport coefficients consistently in their theory.  In our derivation the 
expected result at zero frequency occur naturally in our formalism. In general, we believe the very 
nature of discreteness (as in paramagnetic materials) gives rise to relatively long magnetic relaxation 
times. However, the magnetic relaxation time of a coherent collection of spins (as in FMR) implies shorter 
relaxation times, since it involves collective acoustic waves in the interaction scheme. 
Our present theoretical treatment takes this into account via the continuum magneto-mechanics. 

\section{Theoretical Model \label{tm}}
From Eq.(\ref{intro3}), it is evident that the heating rate per unit volume due to the dissipative 
magnetic intensity \begin{math} {\bf H}^\prime   \end{math} obeys 
\begin{eqnarray}
\dot{Q}=\frac{d{\bf M}}{dt}\cdot {\bf H}^\prime 
\nonumber \\ 
\dot{Q}=
\left[\frac{1}{\gamma M}\right]\frac{d{\bf M}}{dt}\cdot \tilde{\alpha}\cdot \frac{d{\bf M}}{dt}
\nonumber \\ 
\dot{Q}=
\frac{M}{\gamma }\dot{N}_i\alpha_{ij}\dot{N}_j
\ \ {\rm wherein}\ \ {\bf N}=\frac{\bf M}{M}\ ,
 \label{tm1}
\end{eqnarray}
and \begin{math} \tilde{\alpha }  \end{math} is a second rank tensor 
\begin{equation}
\tilde{\alpha } =
\pmatrix{\alpha_{xx} & \alpha_{xy} & \alpha_{xz} \cr
\alpha_{yx} & \alpha_{yy} & \alpha_{yz} \cr
\alpha_{zx} & \alpha_{zy} & \alpha_{zz} }.
\label{tm2}
\end{equation}
The crystal displacement \begin{math} {\bf u} \end{math} yields in elasticity theory \cite{Landau:1986}
the strain tensor 
\begin{equation}
{\sf e}_{ij}=\frac{1}{2}(\partial_iu_j+\partial_ju_i)
\label{tm3}
\end{equation}
In virtue of the magneto-elastic effect \cite{Landauprime:2000}, 
a changing magnetization 
\begin{math} d{\bf M}/dt \end{math} will produce a changing strain 
\begin{math} d{\sf e}/dt \end{math}. In detail, in terms of third rank 
magneto-elastic tensor \begin{math} \Lambda_{ijkl}  \end{math} one finds 
\begin{eqnarray}
{\sf e}_{ij}=\Lambda_{ijkl}N_kN_l, 
\nonumber \\ 
\dot{\sf e}_{ij}=2\lambda_{ijkl}{N}_k\dot{N}_l.
\label{tm4}
\end{eqnarray}
Finally, the fourth rank crystal viscosity tensor, \begin{math} \eta_{ijkl} \end{math} 
dtermines the heating rate per unit volume due to the time dependent strain 
\begin{equation}
\dot{Q}=\dot{\sf e}_{ij}\eta_{ijkl}\dot{\sf e}_{kl}.
\label{tm5}
\end{equation}
Employing Eqs.(\ref{tm4}) and (\ref{tm5}) and comparing the result to Eq.(\ref{tm1}) 
yields the central result of our model. 

For any crystal symmetry the Gilbert damping tensor due to magnetostriction coupling 
is rigorously given by 
\begin{equation}
\alpha_{ij}=
\left[\frac{4\gamma }{M}\right](\Lambda_{nmpi}N_p)\eta_{nmrl}(\Lambda_{rlqj}N_q).
\label{tm6}
\end{equation} 
The following properties of the Gilbert damping tensor Eq.(\ref{tm6}) are worthy of note: 
(i) The Gilbert damping tensor \begin{math} \tilde{\alpha}  \end{math} is inversely proportional 
to the magnetization magnitude \begin{math} M  \end{math}. 
(ii) The Gilbert damping tensor \begin{math} \tilde{\alpha} \end{math} is proportional to the 
squares of the magnetostriction tensor elements. 
(iii) The tensor nature of \begin{math} \tilde{\alpha }  \end{math} dictates that the magnetic 
relaxation is {\em anisotropic}. To a sufficient degree of accuracy, one may employ 
an average of the form 
\begin{equation}
\alpha =\frac{1}{3}tr\{\tilde{\alpha } \}
=\left[\frac{\alpha_{xx}+\alpha_{yy}+\alpha_{zz}}{3}\right]
\label{tm7}
\end{equation}
defining a scalar function \begin{math} \alpha  \end{math}.
(iv) The crystal viscosity tensor \begin{math} \eta_{nmrl} \end{math}
may be employed to describe the acoustic wave damping \cite{Landauprime:1986}. 
For a mode label \begin{math} a \end{math}, e.g. a longitudinal 
(\begin{math} a=L \end{math}) or a transverse (\begin{math} a=T \end{math}) mode,  
the acoustic absorption coefficient at frequency \begin{math} \omega  \end{math} 
is given by \cite{Landauprime:1986} 
\begin{equation}
\tau_a^{-1} =\frac{\omega^2 \eta_a}{2\rho v_a^2},
\label{tm8}
\end{equation}
wherein \begin{math} v_a  \end{math} is the acoustic mode velocity and 
\begin{math} \rho \end{math} is the mass density. Finally, for a cubic crystal, 
there are only two {\em independent} magneto-elastic coefficients which may be defined 
\begin{equation}
\Lambda_{xxxx}=\frac{3}{2}\lambda_{100}
\ \ {\rm and} \ \ \Lambda_{xyxy}=\frac{3}{2}\lambda_{111} 
\label{tm9}
\end{equation}
wherein the Cuachy three index magneto-elastic coefficients are 
\begin{math} \ \lambda_{ijk} \  \end{math}.

\section{Comparison with Experiment \label{CE}} 

\begin{table*}
\caption{\label{tab:table1}Calculated and measured Gilbert damping 
($\alpha$) parameters}
\begin{ruledtabular}
\begin{tabular}{ccccccccccc}
 &$q_{T}$&$\lambda_{100}$&$\lambda_{111}$&$M$&$A$&$\Delta H$&$f$&$\tau$&$\alpha_{\rm th}$
 &$\alpha_{\rm exp}$\\
 Materials&($10^{-6} \rm cm^{-1}$)&($10^{-6}$)&($10^{-6}$)&(G/4$\pi$)&($10^{-6} \rm erg/cm$)
 &(\textit{Oe})&(GHz)&($10^{-13} {\rm sec}$)&($10^{-5}$)&($10^{-5}$)\\ \hline\\
 $\rm Y_{3}Fe_{5}O_{12}$ \footnotemark[1]
 &3.8&1.25&2.8&139&0.40&0.33&9.53&4.4&5.56&9.0 \\
 $\rm Y_{3}Fe_{4}GaO_{12}$ \footnotemark[1]
 &1.46&$-$ 1&$-$ 1&36&0.28&3.0&9.53&4.4&51&76 \\ \\
 
 $\rm Li_{0.5}Fe_{2.5}O_{4}$ \footnotemark[2]
 &8.6&$-$ 8&$+$ 0&310&0.40&2.0&9.50&1.5&26&50 \\
 $\rm NiFe_{2}O_{4}$ \footnotemark[2]
 &7.49&$-$ 63&$-$ 26&270&0.40&35&24.0&710&26&350 \\
 $\rm MgFe_{2}O_{4}$ \footnotemark[2]
 &9.30&$-$ 10&$-$ 1&90&0.1&2.3&4.9&1.5&120&120 \\
 $\rm MnFe_{2}O_{4}$ \footnotemark[2]
 &6.6&$-$ 30&$-$ 5&220&0.4&238&9.2&1.5&930&1040 \\ \\
 
 $\rm BaFe_{12}O_{19}$ \footnotemark[3]
 &9.6& ~ &15&350&0.4&6&55&1.5&18&26 \\ \\
 
 Ni\footnotemark[4]
 &6.3&$-$ 46&25&484&0.75&102&9.53&1.8&770&2600 \\ 
 Fe\footnotemark[4]
 &8.75&20&$-$ 20&1690&1.9&9&9.53&1.8&30&220 \\
 Co\footnotemark[4]
 &5.1& ~ &80&1400&2.78&15&9.53&1.8&530&370 \\  
\end{tabular}
\footnotemark[1]{{\it Garnets}}
\footnotemark[2]{{\it Spinels}}
\footnotemark[3]{{\it Hexagonal Ferrite}}
\footnotemark[4]{{\it Ferromagnetic Materials}}\\
\end{ruledtabular}
{\it (Note: Longitudinal acoustic wave constant is 
\begin{math} q_{L}=(v_{T}/v_{L})q_{T}  \end{math}) }
\end{table*}

The Gilbert damping factor \begin{math} \alpha  \end{math} may be deduced from the measurement 
of the intrinsic FMR linewidth. However, the measurement of the intrinsic linewidth is, indeed, 
very difficult. The reason for this conclusion is that there are too many extrinsic effects that 
influence the measurement. For example, in ferromagnetic metals like Ni, Co and Fe the intrinsic 
linewidth contribution to the total linewidth measurement \cite{Bhagat:1974,Bailey:1972} may be 
between 10\% and 30\%. The rest of the linewidth \cite{Ament:1955} may be due to 
exchange-conductivity effects. 

However, there may be other contributions, such as magnetostatic excitations, surface roughness, 
volume defects \cite{Clogston:1956}, crystal quality, interfaces \cite{Vittoria:1977}, size, etc.. 
Similar conclusions apply to ferrites except there are no exchange-conductivity 
effects \cite{Ament:1955}. Thus, the reader should be mindful that when we quote or cite an 
intrinsic value of the linewidth it represents a maximum value for there can be some hidden 
extrinsic contributions in an experiment. However, we have relied on data well established 
over the years. The criteria that we have adopted in choosing an ensemble of intrinsic 
linewidth measurements are the ones exhibiting the narrowest linewidth ever measured in 
single crystal materials. In addition, we required full knowledge of their elastic, 
magnetic and electrical properties \cite{Bhagat:1974,Bailey:1972,Ament:1955,Landolt:1980}. The
objective is not to introduce any adjustable parameters.

The experimental value of Gilbert damping parameter 
\begin{math} \alpha_{\rm exp} \end{math} may be deduced from the FMR linewidth 
\begin{math} \Delta H  \end{math} at frequency \begin{math} f \end{math} as
\begin{equation}
\alpha_{\rm exp}=\frac{\sqrt{3}}{2}\left(\frac{\gamma \Delta H}{2\pi f}\right).
\label{CE1}
\end{equation} 
The factor \begin{math} \sqrt{3}/2  \end{math} assumes Lorentzian line shape of 
the resonance absorption curve. The theoretical Gilbert damping parameter 
\begin{math} \alpha_{\rm th}  \end{math} value is expressed in terms of known \cite{Bailey:1972} 
parameters so that there are no adjustable parameters in our comparison to 
experiments, as shown in  TABLE \ref{tab:table1}. The theoretical prediction for the Gilbert 
damping paramter is that 
\begin{equation}
\alpha_{\rm th}=\frac{36 \rho \gamma}{M\tau }\left[\frac{\lambda_{100}^2}{q_L^2}
+\frac{\lambda_{111}^2}{q_T^2}\right],
\label{CE2}
\end{equation}
wherein \begin{math} \rho \end{math} is the mass density, 
\begin{math} q_T\approx {\rm v}_T\frac{M}{2\gamma A}  \end{math} is the transverse acoustic 
propagation constant, \begin{math} q_L \end{math} is the longitudinal acoustic propagation 
constant, \begin{math} {\rm v}_T  \end{math} is the transverse sound velocity,
\begin{math} A  \end{math} is the exchange stiffness constant, 
\begin{math} \lambda_{100} \end{math} and \begin{math} \lambda_{111} \end{math} are 
magnetostriction constants for a cubic crystal magnetic material. The transverse acoustic propagation 
constant, was approximated on the basis that the relaxation process conserved energy 
and wave vector. Since the acoustic frequency is fixed in the process the longitudinal 
propagation constant may be also calculated to be 
\begin{math} q_L=q_T({\rm v}_T/{\rm v}_L) \end{math} for magnetic materials, 
wherein \begin{math} {\rm v}_L  \end{math} is the longitudinal sound wave velocity.

In FIG.\ref{Fig1}, we plot the experimental and theoretical values Gilbert damping 
constants as given by Eqs.(\ref{CE1}) and (\ref{CE2}). We note that the agreement 
between theory and experiment is remarkable in view of the fact that any of the cited 
parameters could differ from the ones listed in TABLE \ref{tab:table1} by as much 
as 20-30\%. For example, the linewidth reported in TABLE \ref{tab:table1} may not 
be on the same sample where the elastic or magnetic parameters were cited. In a few 
cases we needed to extrapolate the value of \begin{math} A \end{math}, since there 
was no published value. In FIG.\ref{Fig1}, we did not present data on the ferromagnetic 
metals for lack of confidence on the linewidth data. For example, magnetostatic mode 
excitations have a deleterious effect on the dependence of the FMR linewidth on size. 
Most, if not all, previous FMR linewidth measurements have been performed on slabs, wiskers, etc..
whcih can indeed support magnetostatic mode excitations. Additional complications arise 
as a result of exchange-conductivity excitations in the linewidth data. Nevertheless, 
the agreement between theory and experiment is quite satisfctory.

\section{Conclusion}
\begin{figure}[tp]
\scalebox {0.8}{\includegraphics{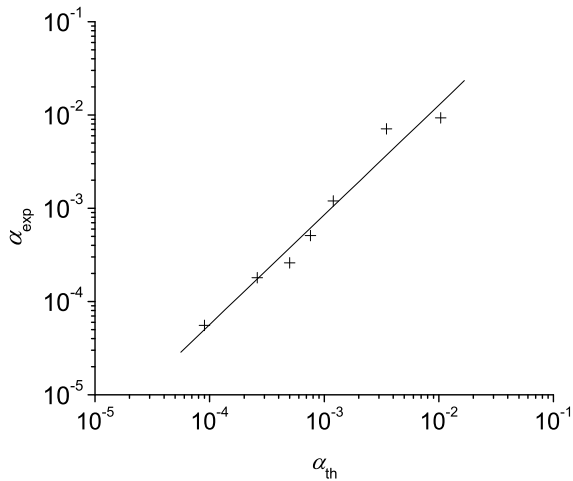}}
\caption{Shown are the experimental and theoretical values of the Gilbert damping 
constants as given by Eqs.(\ref{CE1}) and (\ref{CE2}).}
\label{Fig1}
\end{figure}

Qualitative and quantitatively our model is in agreement with experimental observations
of the intrisic FMR linewidth reported over the years. Speciafically, experimentally 
the most important characteristics of the intrinsic FMR linewidth, \begin{math} \Delta H \end{math}, 
measured on ordered magnetic materials (metal or insulator) for the past fifty years 
are that \begin{math} \Delta H \end{math} scales with frequency and 
\begin{math} \frac{1}{M} \end{math} \cite{Bhagat:1974,Roschmann:1975,Vittoira:1985}. Indeed, 
these are the predictions of our theory. In addition, \begin{math} \Delta H \end{math} scales 
with the magnetostriction constant squared, see FIG.\ref{Fig1}. FIG.\ref{Fig1} was plotted 
in a logarithmic scale only to be able to include all of the data in TABLE \ref{tab:table1}.
\normalfont
Another prediction of our theoretical work is that the Gilbert damping parameter 
\begin{math} \tilde{\alpha} \end{math} is not simply a scalar parameter but a tensor quantity. 
This implies that the FMR linewidth is intrinsically anisotropic in single crystals of ferri-ferromagnetic 
materials. There was much controversy in the seventies about whether or not the intrinsic linewidth 
should be anisotropic or not. Poor quality of samples seemed to have incited the controversy. 
Improved or more accurate angular linewidth data \cite{Ament:1955,Clogston:1956} supports the notion 
of an anisotropic linewidth in ordered magnetic materials in agreement with our model. In summary, 
we believe that the comparison between theory and experiment is very encouraging in terms of continuing 
this continuum approach to explain intrinsic linewidths in ordered magnetic materials.  

{\bf Acknowledgement}\\
We wish to thank to Prof. V.G. Harris and A. Geiler for stimulus discussions 
about magnetic materials and their relaxation.

\end{document}